\documentclass[12pt,preprint]{aastex}

\begin{document}

\title{Infrared Counterparts to {\it Chandra} X-Ray Sources in the
Antennae}

\author{D.M. Clark \altaffilmark{1},  S.S. Eikenberry 
\altaffilmark{1}, B.R. Brandl \altaffilmark{2}, J.C. Wilson 
\altaffilmark{3}, J.C. Carson \altaffilmark{4}, C.P. Henderson 
\altaffilmark{5}, T.L. Hayward \altaffilmark{6}, D.J. Barry 
\altaffilmark{5}, A.F. Ptak \altaffilmark{7}, and E.J.M. Colbert \altaffilmark{7}}

\altaffiltext{1}{Department of Astronomy, University of Florida,
  P.O. Box 112055, Gainesville, FL 32611; dmclark@astro.ufl.edu.}
\altaffiltext{2}{Leiden University, P.O. Box 9513, 2300 RA Leiden, Netherlands.}
\altaffiltext{3}{Department of Astronomy, P.O. Box 400325, University of
Virginia, Charlottesville, VA 22904.}
\altaffiltext{4}{Jet Propulsion Laboratory, Earth and Space Science, Pasadena, CA 91109.}
\altaffiltext{5}{Department of Astronomy, Cornell University, 610
  Space Sciences Building, Ithaca, NY
14853-6801.}
\altaffiltext{6}{Gemini Observatory, AURA, Inc./Casilla 603, La Serena,
Chile.}
\altaffiltext{7}{Department of Physics and Astronomy, Johns Hopkins
University, 3400 North Charles Street, Baltimore, MD 21218.}

\begin{abstract}

    We use deep $J$ ($1.25 \ \mu$m) and $K_s$ ($2.15 \ \mu$m) images
of the Antennae (NGC 4038/9) obtained with the Wide-field InfraRed
Camera on the Palomar 200-inch telescope, together with the {\it
Chandra} X-ray source list of \citet{zez02a}, to search for infrared
counterparts to X-ray point sources.  We establish an X-ray/IR
astrometric frame tie with $\sim0\farcs5$ rms residuals over a $\sim
4\farcm3$ field.  We find 13 ``strong'' IR counterparts brighter than
$K_s = 17.8$ mag and $<1\farcs0$ from X-ray sources, and an additional
6 ``possible'' IR counterparts between $1\farcs0$ and $1\farcs5$ from
X-ray sources.  Based on a detailed study of the surface density of IR
sources near the X-ray sources, we expect only $\sim 2$ of the
``strong'' counterparts and $\sim 3$ of the ``possible'' counterparts
to be chance superpositions of unrelated objects.

    Comparing both strong and possible IR counterparts to our
photometric study of $\sim 220$ IR clusters in the Antennae, we find
with a $>$ 99.9\% confidence level that IR counterparts to X-ray
sources are $\Delta M_{K_s} \sim 1.2$ mag more luminous than average
non-X-ray clusters.  We also note that the X-ray/IR matches are
concentrated in the spiral arms and ``overlap'' regions of the
Antennae.  This implies that these X-ray sources lie in the most
``super'' of the Antennae's Super Star Clusters, and thus trace the
recent massive star formation history here.  Based on the $N_H$
inferred from the X-ray sources without IR counterparts, we determine
that the absence of most of the ``missing'' IR counterparts../../../ASTRO-PH/ is not due
to extinction, but that these sources are intrinsically less luminous
in the IR, implying that they trace a different (possibly older)
stellar population.  We find no clear correlation between X-ray
luminosity classes and IR properties of the sources, though
small-number statistics hamper this analysis.

\end{abstract}


\section{Introduction}

Recently, high resolution X-ray images using {\it Chandra} have
revealed 49 point sources in the Antennae \citep{zez02a}.  We will
assume a distance to the Antennae of 19.3 Mpc (for $H_{0}$=75 km
s$^{-1}$ Mpc$^{-1}$), which implies 10 sources have X-ray luminosities
greater than $10^{39}$ ergs s$^{-1}$.  Considering new observations of
red giant stars in the Antennae indicate a distance of 13.8 Mpc
\citep{sav04}, we point out this ultraluminous X-ray source population
could decrease by roughly a half.  Typically, masses of black holes
produced from standard stellar evolution are less than $\sim20$ 
$M_{\odot}$ \citep[e.g.,][]{fry01}.  The Eddington luminosity limit
implies that X-ray luminosities $>10^{39}$ ergs s$^{-1}$ correspond to
higher-mass objects not formed from a typical star.  Several authors
\citep[e.g.,][]{fab89,zez99,rob00,mak00} suggest these massive ($10
\sbond 1000$ $M_{\sun}$) compact sources outside galactic nuclei are
intermediate mass black holes (IMBHs), a new class of BHs.  While
IMBHs could potentially explain the observed high luminosities, other
theories exist as well, including beamed radiation from a stellar mass
BH \citep{kin01}; super-Eddington accretion onto lower-mass objects
\citep[e.g.,][]{moo03,beg02}; or supernovae exploding in dense
environments \citep{ple95,fab96}.

Compact objects tend to be associated with massive star formation,
which is strongly suspected to be concentrated in young stellar
clusters \citep{lad03}.  Massive stars usually end their lives in
supernovae, producing a compact remnant.  This remnant can be kicked
out of the cluster due to dynamical interactions, stay behind after
the cluster evaporates, or remain embedded in its central regions.
This last case is of particular interest to us as the compact object
is still {\it in situ}, allowing us to investigate its origins via the
ambient cluster population.  The potential for finding such
associations is large in the Antennae due to large numbers of both
X-ray point sources and super star clusters; a further incentive for
studying these galaxies.

In \citet[henceforth Paper I]{bra05} we presented $J$ and $K_s$
photometry of $\sim220$ clusters in the Antennae.  Analysis of ($J -
K_s$) colors indicated that many clusters in the overlap region suffer
from 9--10 mag of extinction in the $V$-band.  This result contrasts
with previous work by \citet{whi02} who associated optical sources
with radio counterparts in the Antennae \citep{nef00} and argued that
extinction is not large in this system.  Here, we continue our
analysis of these Antennae IR images by making a frame-tie between the
IR and {\it Chandra} X-ray images from \citet{fab01}.  Utilizing the
similar dust-penetrating properties of these wavelengths, we
demonstrate the power of this approach to finding counterparts to
X-ray sources.  By comparing the photometric properties of clusters
with and without X-ray counterparts, we seek to understand the cluster
environments of these X-ray sources.  In \S2 we discuss the IR
observations of the Antennae. \S3 explains our matching technique and
the photometric properties of the IR counterparts.  We conclude with a
summary of our results in \S4.

\section{Observations and Data Analysis}

\subsection{Infrared Imaging}

We obtained near-infrared images of NGC 4038/9 on 2002 March 22 using
the Wide-field InfraRed Camera (WIRC) on the Palomar 5-m telescope.
At the time of these observations, WIRC had been commissioned with an
under-sized HAWAII-1 array (prior to installation of the full-sized
HAWAII-2 array in September 2002), providing a $\sim 4.7 \times
4.7$-arcminute field of view with $\sim0\farcs25$ pixels (``WIRC-1K''
-- see \citet{wil03} for details).  Conditions were non-photometric
due to patches of cloud passing through.  Typical seeing-limited
images had stellar full-width at half-maximum of $1\farcs0$ in $K_s$
and $1\farcs3$ in $J$.  We obtained images in both the $J$- ($1.25
\mu$m) and $K_s$-band ($2.15 \mu$m) independently.  The details of the
processing used to obtain the final images are given in Paper 1.

\subsection{Astrometric Frame Ties}

The relative astrometry between the X-ray sources in NGC 4038/9 and
images at other wavelengths is crucial for successful identification
of multi-wavelength counterparts.  Previous attempts at this have
suffered from the crowded nature of the field and confusion between
potential counterparts \citet{zez02b}.  However, the infrared waveband
offers much better hopes for resolving this issue, due to the similar
dust-penetrating properties of photons in the {\it Chandra} and $K_s$
bands.  (See also \citet{bra05} for a comparison of IR extinction to
the previous optical/radio extinction work of \citet{whi02}.)  We thus
proceeded using the infrared images to establish an astrometric
frame-tie, i.e. matching {\it Chandra} coordinates to IR pixel
positions.

As demonstrated by \citet{bau00}, we must take care when searching for
X-ray source counterparts in crowded regions such as the Antennae.
Therefore, our astrometric frame-tie used a unique approach based on
solving a two-dimensional linear mapping function relating right
ascension and declination coordinates in one image with x and y pixel
positions in a second image.  The solution is of the form:

\begin{eqnarray}
r_1 = ax_1+by_1+c, \\
d_1 = dx_1+dy_1+f
\end{eqnarray}

Here $r_1$ and $d_1$ are the right ascension and declination,
respectively, for a single source in one frame corresponding to the
$x_1$ and $y_1$ pixel positions in another frame.  This function
considers both the offset and rotation between each frame.  Since we
are interested in solving for the coefficients $a$--$f$, elementary
linear algebra indicates we need six equations or three separate
matches.  Therefore, we need at least three matches to fully describe
the rms positional uncertainty of the frame-tie.

We first used the above method to derive an approximate astrometric
solution for the WIRC $K_s$ image utilizing the presence of six
relatively bright, compact IR sources which are also present in images
from the 2-Micron All-Sky Survey (2MASS).  We calculated pixel
centroids of these objects in both the 2MASS and WIRC images, and used
the 2MASS astrometric header information to convert the 2MASS pixel
centroids into RA and Dec.  These sources are listed in Table 1.
Applying these six matches to our fitting function we found a small
rms positional uncertainty of $0\farcs2$, which demonstrates an
accurate frame-tie between the 2MASS and WIRC images.

Using the 2MASS astrometric solution as a baseline, we identified
seven clear matches between {\it Chandra} and WIRC sources, which had
bright compact IR counterparts with no potentially confusing sources
nearby (listed in Table 2).  We then applied the procedure described
above, using the {\it Chandra} coordinates listed in Table 1 of
\citet{zez02a} (see that reference for details on the {\it Chandra}
astrometry) and the WIRC pixel centroids, and derived the astrometric
solution for the IR images in the X-ray coordinate frame.  For the 7
matches, we find an rms residual positional uncertainty of $\sim
0\farcs5$ which we adopt as our $1 \sigma$ position uncertainty.  We
note that the positional uncertainty is an entirely {\it empirical}
quantity.  It shows the achieved uncertainty in mapping a target from
one image reference frame to the reference frame in another band, and
automatically incorporates all contributing sources of uncertainty in
it.  These include, but are not limited to, systematic uncertainties
(i.e. field distortion, PSF variation, etc. in both {\it Chandra} and
WIRC) and random uncertainties (i.e. centroid shifts induced by photon
noise, flatfield noise, etc. in both {\it Chandra} and WIRC).  Thus,
given the empirical nature of this uncertainty, we expect it to
provide a robust measure of the actual mapping error -- an expectation
which seems to be borne out by the counterpart identification in the
following section.

To further test the accuracy of our astrometric solution we explored the range
in rms positional uncertainties for several different frame-ties.
Specifically we picked ten IR/X-ray matches separated by
$<$1$\arcsec$, which are listed in Table 3 (see \S3.1).  Of these ten
we chose 24 different combinations of seven matches resulting in 24
unique frame-ties.  Computing the rms positional uncertainty for each
we found a mean of 0$\farcs$4 with a 1$\sigma$ uncertainty of
0$\farcs$1.  Considering the rms positional uncertainty for the
frame-tie used in our analysis falls within 1$\sigma$ of this mean
rms, this indicates we made an accurate astrometric match between the
IR and X-ray.

\subsection{Infrared Photometry}

We performed aperture photometry on 222 clusters in the $J$-band and
221 clusters in the $K_s$-band (see also Paper 1).  We found that the
full width at half maximum (FWHM) was 3.5 pixels ($0\farcs9$) in the
$K_s$ image and 4.6 pixels ($1\farcs2$) in the $J$ band.  We used a
photometric aperture of 5-pixel radius in $K_s$ band, and 6-pixel
radius in $J$ band, corresponding to $\sim 3\sigma$ of the Gaussian
PSF.

Background subtraction is both very important and very difficult in an
environment such as the Antennae due to the brightness and complex
structure of the underlying galaxies and the plethora of nearby
clusters.  In order to address the uncertainties in background
subtraction, we measured the background in two separate annuli around
each source: one from 9 to 12 pixels and another from 12 to 15 pixels.
Due to the high concentration of clusters, crowding became an issue.
To circumvent this problem, we employed the use of sky background arcs
instead of annuli for some sources.  These were defined by a position
angle and opening angle with respect to the source center.  All radii
were kept constant to ensure consistency.  In addition, nearby bright
sources could shift the computed central peak position by as much as a
pixel or two.  If the centroid position determined for a given source
differed significantly ($>1$ pixel) from the apparent brightness peak
due to such contamination, we forced the center of all photometric
apertures to be at the apparent brightness peak.  For both annular
regions, we calculated the mean and median backgrounds per pixel.

Multiplying these by the area of the central aperture, these values
were subtracted from the flux measurement of the central aperture to
yield 4 flux values for the source in terms of DN.  Averaging these
four values provided us with a flux value for each cluster.  We
computed errors by considering both variations in sky background,
$\sigma_{sky}$, and Poisson noise, $\sigma_{adu}$.  We computed
$\sigma_{sky}$ by taking the standard deviation of the four measured
flux values.  We then calculated the expected Poisson noise by scaling
DN to $e^-$ using the known gain of WIRC \citep[2 $e^{-1}$
DN$^{-1}$,][]{wil03} and taking the square root of this value.  We
added both terms in quadrature to find the total estimated error in
photometry.

We then calibrated our photometry using the bright 2MASS star in Table 1.

\section{Results and Discussion}

\subsection{Identification of IR Counterparts to {\it Chandra} Sources}

We used the astrometric frame-tie described above to identify IR
counterpart candidates to {\it Chandra} X-ray sources in the WIRC
$K_s$ image.  We restricted our analysis to sources brighter than
$K_s\sim19.4$ mag.  This is our $K_s$ sensitivity limit which we
define in our photometric analysis below (see \S3.2.1).  Using the
$0\farcs5$ rms positional uncertainty of our frame-tie, we defined
circles with 2$\sigma$ and 3$\sigma$ radii around each {\it Chandra}
X-ray source where we searched for IR counterparts (Figure 1).  If an
IR source lay within a $1\farcs0$ radius ($2\sigma$) of an X-ray
source, we labeled these counterparts as ``strong''.  Those IR sources
between $1\farcs0$ and $1\farcs5$ ($2-3 \sigma$) from an X-ray source
we labeled as ``possible'' counterparts.  We found a total of 13
strong and 6 possible counterparts to X-ray sources in the Antennae.
These sources are listed in Table 3 and shown in Figure 3.  Of the 19
X-ray sources with counterparts, two are the nuclei \citep{zez02a},
one is a background quasar \citep{cla05}, and two share the same IR
counterpart.  Therefore, in our analysis of cluster properties, we
only consider the 15 IR counterparts that are clusters.  (While
X-42 has two IR counterparts, we chose the closer, fainter cluster for
our analysis.)

We then attempted to estimate the level of ``contamination'' of these
samples due to chance superposition of unrelated X-ray sources and IR
clusters.  This estimation can be significantly complicated by the
complex structure and non-uniform distribution of both X-ray sources
and IR clusters in the Antennae, so we developed a simple, practical
approach.  Given the $<0\farcs5$ rms residuals in our relative
astrometry for sources in Table 2, we assume that any IR clusters
lying in a background annulus with radial size of
$2\farcs0$--$3\farcs0$ ($4-6\sigma$) centered on all X-ray source
positions are chance alignments, with no real physical connection (see
Figure 1).  Dividing the total number of IR sources within the
background annuli of the 49 X-ray source positions by the total area
of these annuli, we find a background IR source surface density of
0.02 arcsecond$^{-2}$ near {\it Chandra} X-ray sources.  Multiplying
this surface density by the total area of all ``strong'' regions
($1\farcs0$ radius circles) and ``possible'' regions ($1\farcs0$ --
$1\farcs5$ annuli) around the 49 X-ray source positions, we estimated
the level of source contamination contributing to our ``strong'' and
``possible'' IR counterpart candidates.  We expect two with a
$1\sigma$ uncertainty of +0.2/-0.1\footnote{Found using confidence
levels for small number statistics listed in Tables 1 and 2 of
\citet{geh86}.} of the 13 ``strong'' counterparts to be due to chance
superpositions, and three with a $1\sigma$ uncertainty of
+0.5/-0.3\footnotemark[\value{footnote}] of the six ``possible''
counterparts to be chance superpositions.

This result has several important implications.  First of all, it is
clear that we have a significant excess of IR counterparts within
$1\farcs0$ of the X-ray sources -- 13, where we expect only two in the
null hypothesis of no physical counterparts.  Even including the
``possible'' counterparts out to $1\farcs5$, we have a total of 19
counterparts, where we expect only five from chance superposition.
Secondly, this implies that for any given ``strong'' IR counterpart,
we have a probability of $\sim$ 85\% ($11/13$ with a $1\sigma$
uncertainty of 0.3\footnotemark[\value{footnote}]) that the
association with an X-ray source is real.  Even for the ``possible''
counterparts, the probability of true association is $\sim$50\%. These
levels of certainty are a tremendous improvement over the
X-ray/optical associations provided by \citet{zez02b}, and are strong
motivators for follow-up multi-wavelength studies of the IR
counterparts.  Finally, we can also conclude from strong concentration
of IR counterparts within $\sim 1 \arcsec$ of X-ray sources that the
frame tie uncertainty estimates described above are reasonable.

Figure 2 is a $4\farcm3\times4\farcm3$ $K_s$ image of the Antennae
with X-ray source positions overlaid.  We designate those X-ray
sources with counterparts using red circles.  Notice that those
sources with counterparts lie in the spiral arms and bridge region of
the Antennae.  Since these regions are abundant in star formation,
this seems to indicate many of the X-ray sources in the Antennae are
tied to star formation in these galaxies.

\subsection{Photometric Properties of the IR Counterparts}

\subsubsection{Color Magnitude Diagrams}

Using the 219 clusters that had both $J$ and $K_s$ photometry, we made
$(J-K_s)$ versus $K_s$ color magnitude diagrams (Figure 4).  We
estimated our sensitivity limit by first finding all clusters with
signal-to-noise $\sim5\sigma$.  The mean $J$ and $K_s$ magnitudes for
these clusters were computed separately and defined as cutoff values
for statistical analyzes.  This yielded 19.0 mag in $J$ and 19.4 mag
in $K_s$.  We note that the X-ray clusters are generally bright in the
IR compared to the general population of clusters.  While the IR
counterpart for one X-ray source (X-32) falls below our $J$-band
sensitivity limit, its $K_s$ magnitude is still above our $K_s$
cutoff.  Therefore, we retained this source in our analysis.

We then broke down the X-ray sources into three luminosity classes
(Figure 4).  We took the absorption-corrected X-ray luminosities,
$L_X$, as listed in Table 1 of \citet{zez02a} for all sources of
interest.  These luminosities assumed a distance to the Antennae of 29
Mpc.  We used 19.3 Mpc (for $H_{0}$=75 km s$^{-1}$ Mpc$^{-1}$) instead
and so divided these values by 2.25 as suggested in \citet{zez02a}.
We defined the three X-ray luminosities as follows: Low Luminosity
X-ray sources (LLX's) had $L_{X}$ $<$ 3$\times$$10^{38}$ ergs
s$^{-1}$, High Luminosity X-ray sources (HLX's) were between $L_{X}$
of 3$\times10^{38}$ergs s$^{-1}$ and 1$\times10^{39}$ ergs
s$^{-1}$, while $L_{X}$ $>$ 1$\times10^{39}$ ergs s$^{-1}$ were
Ultra-Luminous X-ray Sources (ULX's).  In Figure 4 we designate each
IR counterpart according to the luminosity class of its corresponding
X-ray source.  There does not appear to be a noticeable trend in the
IR cluster counterparts between these different groupings.

\subsubsection{Absolute K Magnitudes}

To further study the properties of these IR counterparts, we
calculated $M_{K_s}$ for all IR clusters.  We calculated reddening
using the observed colors, $(J-K_s)_{obs}$, (hence forth the ``color
method'').  Assuming all clusters are dominated by O and B stars,
their intrinsic $(J-K_s)$ colors are $\sim$0.2 mag.  Approximating
this value as 0 mag, this allowed us to estimate $A_{K_s}$ as $\simeq$
$(J-K_s)_{obs}$/1.33 using the extinction law defined in
\citet[hereafter CCM]{car89}.  Since these derived reddenings are
biased towards young clusters, they will lead to an overestimate of
$M_{K_s}$ for older clusters.

For IR counterparts to X-ray sources, we also computed X-ray-estimated
$A_{K_s}$ using the column densities, $N_{H}$, listed in Table 5 of
\citet{zez02a}.  Here, $N_{H}$ is derived by fitting both a Power Law
(PL) and Raymond-Smith \citep[RS][]{ray77} model to the X-ray spectra.
Using the CCM law, $A_{K_s}$ is defined as 0.12$A_{V}$.  Taking
$A_{V}$ = $5\times10^{-22}$ mag cm$^2$ $N_{H}$, we could then derive
$A_{K_s}$.

Then we compared $A_{K_s}$ calculated using the ``color method'' to
$A_{K_s}$ found using the above two $N_{H}$ models.  We found the
``color method'' matched closest to $N_{H}(PL)$ for all except one
(the cluster associated with {\it Chandra} source 32 as designated in
\citet{zez02b}).

In Figure 5, we plot histograms of the distribution of $K_s$-band
luminosity, $M_{K_s}$.  Figure 5 displays all clusters as well as over
plotting only those with X-ray counterparts.  Notice that the clusters
with associated X-ray sources look more luminous.  To study whether
this apparent trend in luminosity is real, we compared these two
distributions using a two-sided Kolmogorov-Smirnov (KS) test.  In our
analysis, we only included clusters below $M_{K_s}$ $<$ -13.2 mag.
Restricting our study to sources with ``good'' photometry, we first
defined a limit in $K_s$, 18.2 mag, using the limiting $J$ magnitude,
19.0 mag as stated above, and, since the limit in $K_s$ is a function
of cluster color, the median $(J-K_s)$ of 0.8 mag.  Subtracting the
distance modulus to the Antennae, 31.4 mag, from this $K_s$ limit, we
computed our cutoff in $M_{K_s}$.  Since all clusters with X-ray
sources fall below this cutoff, our subsample is sufficient to perform
a statistical comparison.

The KS test yielded a D-statistic of 0.37 with a probability of
$3.2\times10^{-2}$ that the two distributions of clusters with and
without associated X-ray sources are related.  Considering the
separate cluster populations as two probability distributions, each
can be expressed as a cumulative distribution.  The D-statistic is
then the absolute value of the maximum difference between each
cumulative distribution.  This test indicates that those clusters with
X-ray counterparts are more luminous than most clusters in the
Antennae.

\subsubsection{Cluster Mass Estimates}

\citet{whi99} found 70\% of the bright clusters observed with the {\it
Hubble Space Telescope} have ages $<$20 Myr.  Therefore, in this study
we will assume all clusters are typically the same age, $\sim$20 Myr.
This allows us to make the simplifying assumption that cluster mass is
proportional to luminosity and ask: Does the cluster mass affect the
propensity for a given progenitor star to produce an X-ray binary?  We
estimated cluster mass using $K_s$ luminosity ($M_{K_s}$).  Since
cluster mass increases linearly with flux (for an assumed constant age
of all clusters), we converted $M_{K_s}$ to flux.  Using the data as
binned in the $M_{K_s}$ histogram (Figure 5), we calculated an average
flux per bin.  By computing the fraction of number of clusters per
average flux, we are in essence asking what is the probability of
finding a cluster with a specific mass.  Since those clusters with
X-ray sources are more luminous, we expect a higher probability of
finding an X-ray source in a more massive cluster.  As seen in Figure
6, this trend does seem to be true.  Applying a KS-test between the
distributions for all clusters and those associated with X-ray sources
for clusters below the $M_{K_s}$ completeness limit defined in the
previous section, we find a D-statistic of 0.66 and a probability of
$7.2\times10^{-3}$ that they are the same.  Hence, the two
distributions are distinct, indicating it is statistically significant
that more massive clusters tend to contain X-ray sources.

While we assume all clusters are $\sim$20 Myr above, we note that the
actual range in ages is $\sim$1--100 Myr \citep{whi99}.
Bruzual-Charlot spectral photometric models \citep{bru03} indicate
that clusters in this age range could vary by a factor of roughly 100
in mass for a given $K_s$ luminosity.  Thus, we emphasize that the
analysis above should be taken as suggestive rather than conclusive
evidence, and note that in a future paper (Clark, et al. 2006, in
preparation) we explore this line of investigation and the impacts of
age variations on the result in depth.

\subsubsection{Non-detections of IR Counterparts to X-ray Sources}

To assess whether our counterpart detections were dependent of
reddening or their intrinsic brightness, we found limiting values for
$M_{K_s}$ for those X-ray sources without detected IR counterparts.
We achieved this by setting all clusters $K_s$ magnitudes equal to our
completeness limit defined for the CMDs (19.4 mag; see \S3.2.1) and
then finding $M_{K_s(lim)}$ for each using $A_{K_s}$ calculated for
that cluster.  Since $M_{K_s(lim)}$ is theoretical and only depends on
reddening, we could now find this limit for all X-ray sources using an
$A_{K_s}$ estimated from the observed $N_{H}$ values.  Thus we
considered all IR counterparts (detections) and those X-ray sources
without a counterpart (nondetections).  If nondetections are due to
reddening there should not exist a difference in $M_{K_s(lim)}$
between detections and nondetections.  In contrast, if nondetections
are intrinsically fainter, we expect a higher $M_{K_s(lim)}$ for these
sources.  In the case of detections, we considered reddening from both
the ``color method'' and the $N_{H}(PL)$ separately.  We could only
derive nondetections using $N_{H}(PL)$ reddening.  Figure 7 shows
$M_{K_s(lim)}$ appears higher for all nondetections. To test if this
observation is significant, we applied a KS-test to investigate
whether detections and nondetections are separate distributions.  We
find a D-statistic of 0.82 and probability of $8.8\times10^{-6}$ that
these two distributions are the same using the ``color method'' for
detections. Considering the $N_{H}(PL)$ reddening method for
detections instead, the D-statistic drops to 0.48 and the probability
increases to $3.9\times10^{-2}$.  Since both tests indicate these
distributions are distinct, the observed high $M_{K_s(lim)}$ for
nondetections seems to be real.  This leads to the conclusion that
these sources were undetected because they are intrinsically IR-faint,
and that reddening does not play the dominant role in nondetections.

We summarize these statistics in Table 4.  Here we calculated the mean
$K_s$, $(J-K_s)$, and $M_{K_s}$ for three different categories: 1) all
clusters, 2) clusters only connected with X-ray sources, and 3) these
clusters broken down by luminosity class.  We also include
uncertainties in each quantity.  Notice that the IR counterparts
appear brighter in $K_s$ and intrinsically more luminous than most
clusters in the Antennae, although there is no significant trend in
color.  We also summarize the above KS-test results in Table 5.

\section{Conclusions}

We have demonstrated a successful method for finding counterparts to
X-ray sources in the Antennae using IR wavelengths.  We mapped {\it
Chandra} X-ray coordinates to WIRC pixel positions with a positional
uncertainty of $\sim 0\farcs5$.  Using this precise frame-tie we
found 13 ``strong'' matches ($< 1\farcs0$ separation) and 5
``possible'' matches ($1 - 1\farcs5$ separation) between X-ray
sources and IR counterparts.  After performing a spatial and
photometric analysis of these counterparts, we reached the following
conclusions:

1.  We expect only 2 of the 13 ``strong'' IR counterparts to be chance
superpositions.  Including all 19 IR counterparts, we estimated 5 are
unrelated associations.  Clearly, a large majority of the X-ray/IR
associations are real.

2.  The IR counterparts tend to reside in the spiral arms and bridge
region between these interacting galaxies.  Since these regions
contain the heaviest amounts of star formation, it seems evident that
many of the X-ray sources are closely tied to star formation in this
pair of galaxies.

3.  A $K_s$ vs. $(J - K_s)$ CMD reveal those clusters associated with
X-ray sources are brighter in $K_s$ but there does not seem to be a
trend in color.  Separating clusters by the X-ray luminosity classes
of their X-ray counterpart does not reveal any significant trends.

4.  Using reddening derived $(J - K_s)$ colors as well as from
X-ray-derived $N_H$, we found $K_s$-band luminosity for all clusters.
A comparison reveals those clusters associated with X-ray sources are
more luminous than most clusters in the Antennae.  A KS-test indicates
a significant difference between X-ray counterpart clusters and the
general population of clusters.

5.  By relating flux to cluster luminosity, simplistically assuming a
constant age for all clusters, we estimated cluster mass.  Computing
the fraction of number of clusters per average flux, we estimated the
probability of finding a cluster with a specific mass.  We find more
massive clusters are more likely to contain X-ray sources, even after
we normalize by mass.

6.  We computed a theoretical, limiting $M_{K_s}$ for all counterparts
to X-ray sources in the Antennae using X-ray-derived reddenings.
Comparing detections to non-detections, we found those clusters with
X-ray source are intrinsically more luminous in the IR.

In a future paper exploring the effects of cluster mass on XRB
formation rate (Clark, et al. 2006a, in preparation), we will
investigate the effects of age on cluster luminosity and hence our
cluster mass estimates.  Another paper will extend our study of the
Antennae to optical wavelengths (Clark, et al. 2006b, in preparation).
Through an in depth, multi-wavelength investigation we hope to achieve
a more complete picture of counterparts to several X-ray sources in
these colliding galaxies.

\acknowledgments

The authors thank the staff of Palomar Observatory for their excellent
assistance in commissioning WIRC and obtaining these data.  WIRC was
made possible by support from the NSF (NSF-AST0328522), the Norris
Foundation, and Cornell University.  S.S.E. and D.M.C. are supported
in part by an NSF CAREER award (NSF-9983830).  We also thank
J.R. Houck for his support of the WIRC instrument project.

\vfill \eject

\begin{deluxetable}{lcccc}
\tablecaption{Common Sources Used for the 2MASS/WIRC Astrometric Frame Tie}
\tablewidth{0pt}
\startdata
\tableline
\tableline
Description & RA (2MASS) & Dec (2MASS) & $J$  \tablenotemark{1} &
$K_s$ \tablenotemark{1} \\
\tableline
Bright Star & 12:01:47.90 & -18:51:15.8 & 13.07(0.01) & 12.77(0.01) \\
Southern Nucleus & 12:01:53.50 & -18:53:10.0 & 13.45(0.05) & 12.50(0.03) \\
Cluster 1 & 12:01:51.66 & -18:51:34.7 & 14.84(0.01) & 14.24(0.01) \\
Cluster 2 & 12:01:50.40 & -18:52:12.2 & 14.98(0.02) & 14.06(0.01) \\
Cluster 3 & 12:01:54.56 & -18:53:04.0 & 15.05(0.03) & 14.27(0.02) \\
Cluster 4 & 12:01:54.95 & -18:53:05.8 & 16.52(0.01) & 14.66(0.01) \\
\enddata
\tablenotetext{1}{Units are magnitudes, from WIRC photometry. Values
in parentheses indicate uncertainties.}
\end{deluxetable}

\begin{deluxetable}{lccc}
\tablecaption{Common Sources Used for the {\it Chandra}/WIRC
Astrometric Frame Tie}
\tablewidth{0pt}
\startdata
\tableline
\tableline
{\it Chandra} Source ID \tablenotemark{1} & RA ({\it Chandra}) & Dec
({\it Chandra}) & $K_s$ \tablenotemark{2} \\ 
\tableline
6 & 12:01:50.51 & $-$18:52:04.80 & 15.66(0.02) \\
10 & 12:01:51.27 & $-$18:51:46.60 & 14.66(0.01) \\
24 & 12:01:52.99 & $-$18:52:03.20 & 13.55(0.11) \\
29 & 12:01:53.49 & $-$18:53:11.10 & 12.50(0.03) \\
34 & 12:01:54.55 & $-$18:53:03.20 & 14.27(0.02) \\
36 & 12:01:54.81 & $-$18:52:14.00 & 15.92(0.01) \\
37 & 12:01:54.98 & $-$18:53:15.10 & 16.16(0.01) \\
\enddata
\tablenotetext{1}{ID numbers follow the naming convention of \citet{zez02a}.}
\tablenotetext{2}{Units are magnitudes, from WIRC photometry. Values
in parentheses indicate uncertainties.}
\end{deluxetable}

\begin{deluxetable}{lcccccc}
\tablecaption{Potential IR Counterparts to {\it Chandra} X-Ray Sources}
\tablewidth{0pt}
\startdata
\tableline
\tableline
 &  &  & $\Delta\alpha$ \tablenotemark{3} & $\Delta \delta$  \tablenotemark{3} &  & 
\\
{\it Chandra} Src ID \tablenotemark{1} & RA \tablenotemark{2} & Dec
\tablenotemark{2} & (arcsec) & (arcsec) & $J$ \tablenotemark{4} & $K_s$ \tablenotemark{4}
\\
\tableline
``Strong'' Counterparts \\
\tableline
 6 & 12:01:50.51 & -18:52:04.77 & 0.29 & 0.04 & 16.21(0.01) & 15.66(0.02) \\
10 & 12:01:51.27 & -18:51:46.58 & 0.24 & 0.31 & 15.57(0.01) & 14.66(0.01) \\
11 & 12:01:51.32 & -18:52:25.46 & 0.34 & 0.03 & 18.27(0.01) & 17.37(0.08) \\
20 & 12:01:52.74 & -18:51:30.06 & 0.11 & 0.38 & 18.48(0.04) & 17.78(0.02) \\
24 & 12:01:52.99 & -18:52:03.18 & 0.07 & 0.82 & 14.37(0.11) & 13.55(0.11) \\
26 & 12:01:53.13 & -18:52:05.53 & 0.27 & 0.87 & 15.95(0.01) & 14.71(0.15) \\
29 & 12:01:53.49 & -18:53:11.08 & 0.20 & 0.25 & 13.45(0.05) & 12.50(0.03) \\
33 & 12:01:54.50 & -18:53:06.82 & 0.11 & 0.99 & 16.71(0.08) & 16.45(0.07) \\
34 & 12:01:54.55 & -18:53:03.23 & 0.02 & 0.39 & 15.05(0.03) & 14.27(0.02) \\
36 & 12:01:54.81 & -18:52:13.99 & 0.11 & 0.50 & 16.60(0.03) & 15.92(0.01) \\
37 & 12:01:54.98 & -18:53:15.07 & 0.13 & 0.10 & 17.55(0.02) & 16.16(0.01) \\
39 & 12:01:55.18 & -18:52:47.50 & 0.18 & 0.03 & 17.10(0.07) & 15.71(0.04) \\
42 & 12:01:55.65 & -18:52:15.06 & 0.73 & 0.40 & 17.13(0.03) & 16.27(0.06) \\
``Possible'' Counterparts \\             
\tableline                               
15 & 12:01:51.98 & -18:52:26.47 & 1.09 & 0.84 & 16.63(0.04) & 15.95(0.02) \\
22 & 12:01:52.89 & -18:52:10.03 & 0.70 & 1.20 & 15.77(0.01) & 15.13(0.06) \\
25 & 12:01:53.00 & -18:52:09.59 & 0.87 & 0.76 & 15.77(0.01) & 15.13(0.06) \\
32 & 12:01:54.35 & -18:52:10.31 & 0.92 & 1.39 & 20.30(0.45) & 16.84(0.03) \\
35 & 12:01:54.77 & -18:52:52.43 & 0.42 & 0.92 & 16.76(0.02) & 14.88(0.04) \\
40 & 12:01:55.38 & -18:52:50.53 & 0.61 & 1.24 & 16.21(0.02) & 15.27(0.04) \\
\enddata
\tablenotetext{1}{ID numbers follow the naming convention of \citet{zez02a}}
\tablenotetext{2}{{\it Chandra} coordinates  with an uncertainty of
  $0\farcs5$ \citep{zez02a}.}
\tablenotetext{3}{Positional offsets in units of seconds of arc from the {\it
Chandra} coordinates to the WIRC coordinates of the proposed
counterpart.}
\tablenotetext{4}{Units are magnitudes, from WIRC photometry.
Values in parentheses indicate uncertainties in the final listed
digit.}
\end{deluxetable}

\begin{deluxetable}{lcccccc}
\tablecaption{Summary of Potential IR Counterpart Properties.}
\tablewidth{0pt}
\startdata
\tableline
\tableline
Category & $\overline{K}$ & $\sigma_{\overline{K}}$\tablenotemark{1} &
$\overline{(J-K)}$
 & $\sigma_{\overline{(J-K)}}$\tablenotemark{1} & $\overline{M_{K_s}}$
& 
$\sigma_{\overline{M_K}}$\tablenotemark{1} \\
\tableline
all clusters & 16.72 & 0.08 & 0.82 & 0.03 & -15.33 & 0.09 \\
X-ray sources & 15.72 & 0.27 & 0.95 & 0.11 & -16.30 & 0.35 \\
LLX & 15.85 & 0.36 & 0.84 & 0.11 & -16.16& 0.39 \\
HLX & 15.09 & 0.37 & 1.13 & 0.28 & -17.14 & 0.54 \\
ULX & 16.82 & 0.55 & 0.88 & 0.02 & -14.67 & 0.51 \\
\enddata
\tablenotetext{1}{Uncertainties in each value.}
\end{deluxetable}

\clearpage

\begin{deluxetable}{lcc}
\tablecaption{Summary of KS-Test Results.}
\tablewidth{0pt}
\startdata
\tableline
\tableline
& Probability\tablenotemark{1}  & D\tablenotemark{2} \\
\tableline
$M_{K_s}$ & $3.2\times10^{-2}$ & 0.37\\
\tableline
Cluster Mass & $9.6\times10^{-2}$ & 0.40 \\
\tableline
$M_{K_s(lim)}$ (CM)\tablenotemark{3} & $8.8\times10^{-6}$ & 0.82 \\
$M_{K_s(lim)}$ (NH)\tablenotemark{3} & $3.9\times10^{-2}$ & 0.48 \\
\enddata
\tablenotetext{1}{Probability distributions are the same.}
\tablenotetext{2}{Kolmogorov-Smirnov D-statistic.}
\tablenotetext{3}{CM: ``color method'' for detections, NH: $N_H$(PL)
method for detections.  See text for details.}
\end{deluxetable}

\clearpage

\begin{figure}
\figurenum{1}
\plotone{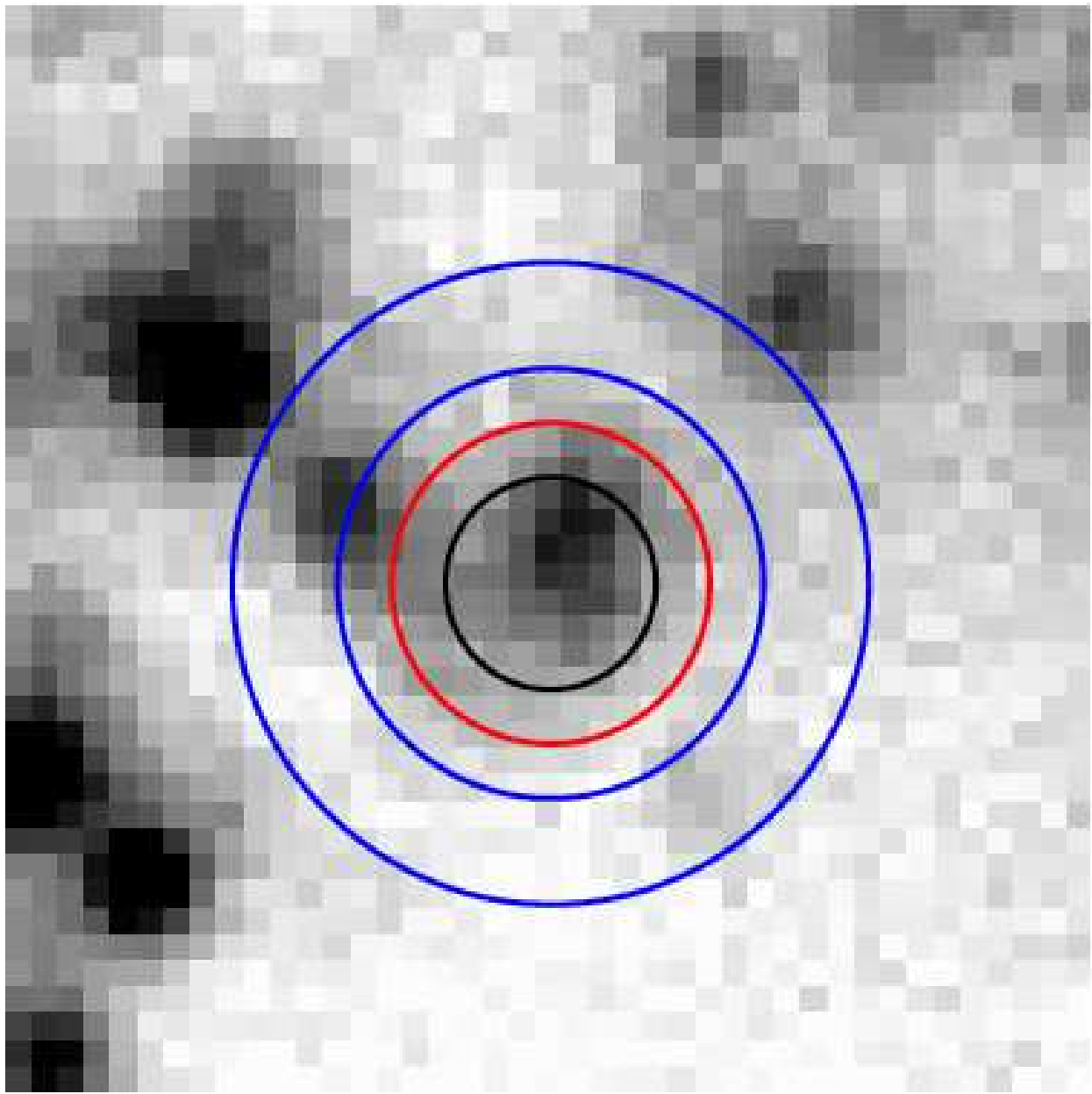}
\caption{IR counterpart to X-36 overlaid with areas of positional
uncertainty centered on the X-ray source position.  Black: 1$\farcs$0
radius circle for strong sources, red: 1$\farcs$0 -- 1$\farcs$5
annulus for possible sources, and blue: 2$\farcs$0 -- 3$\farcs$0
annulus used to estimate background source contamination.
\label{Fig.1}}
\end{figure}

\clearpage

\begin{figure}
\figurenum{2}
\epsscale{1.0}
\plotone{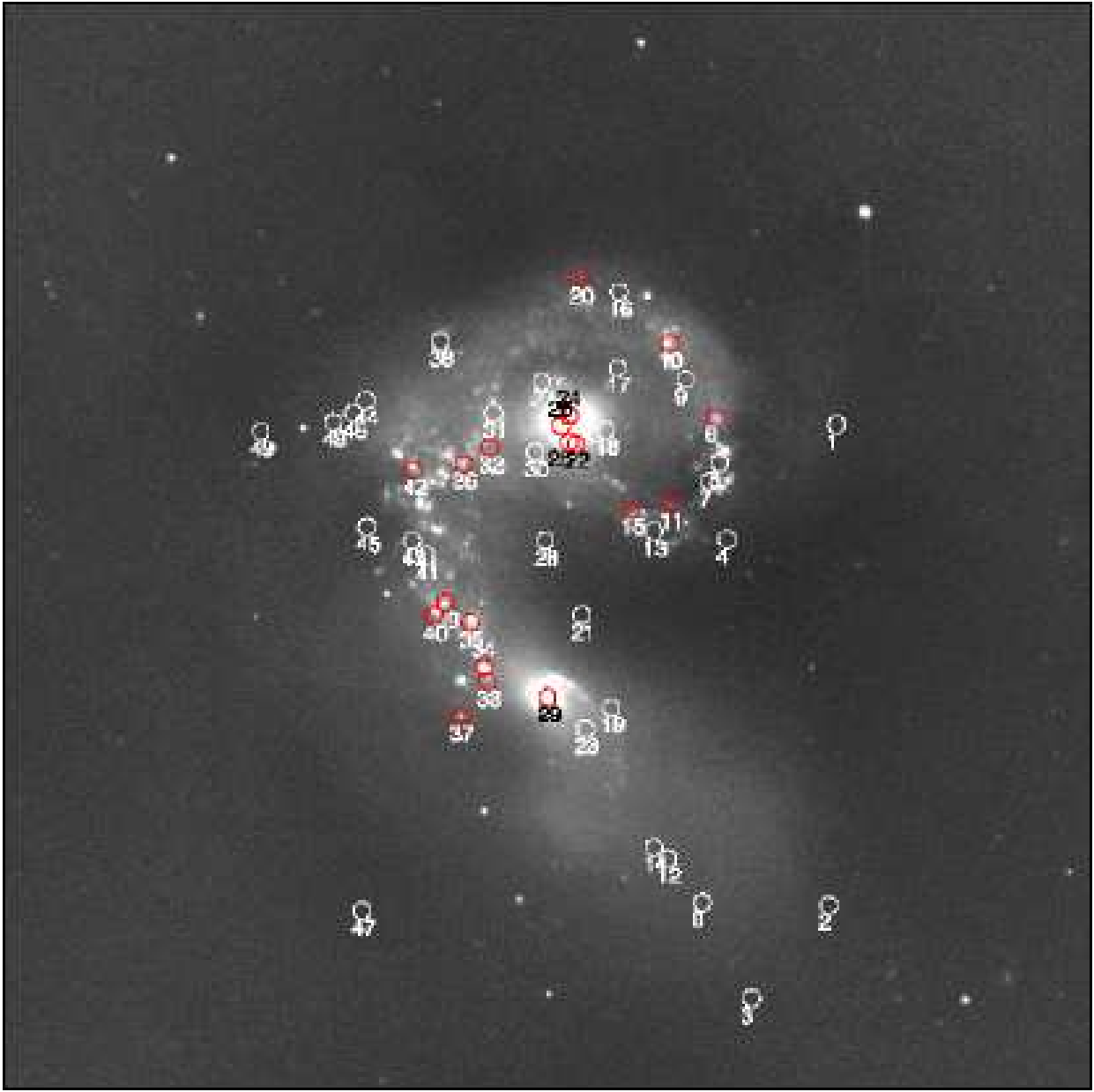}
\caption{$K_s$-band image of the Antennae showing positions of X-ray
sources.  The field-of-view is $4\farcs3\times4\farcs3$.  Red circles
designate those sources with IR counterparts.  Notice these tend to
reside in the spiral arms and bridge region between the galaxies.
\label{Fig.2}}
\end{figure}

\clearpage

\begin{figure}
\epsscale{1.0}
\plottwo{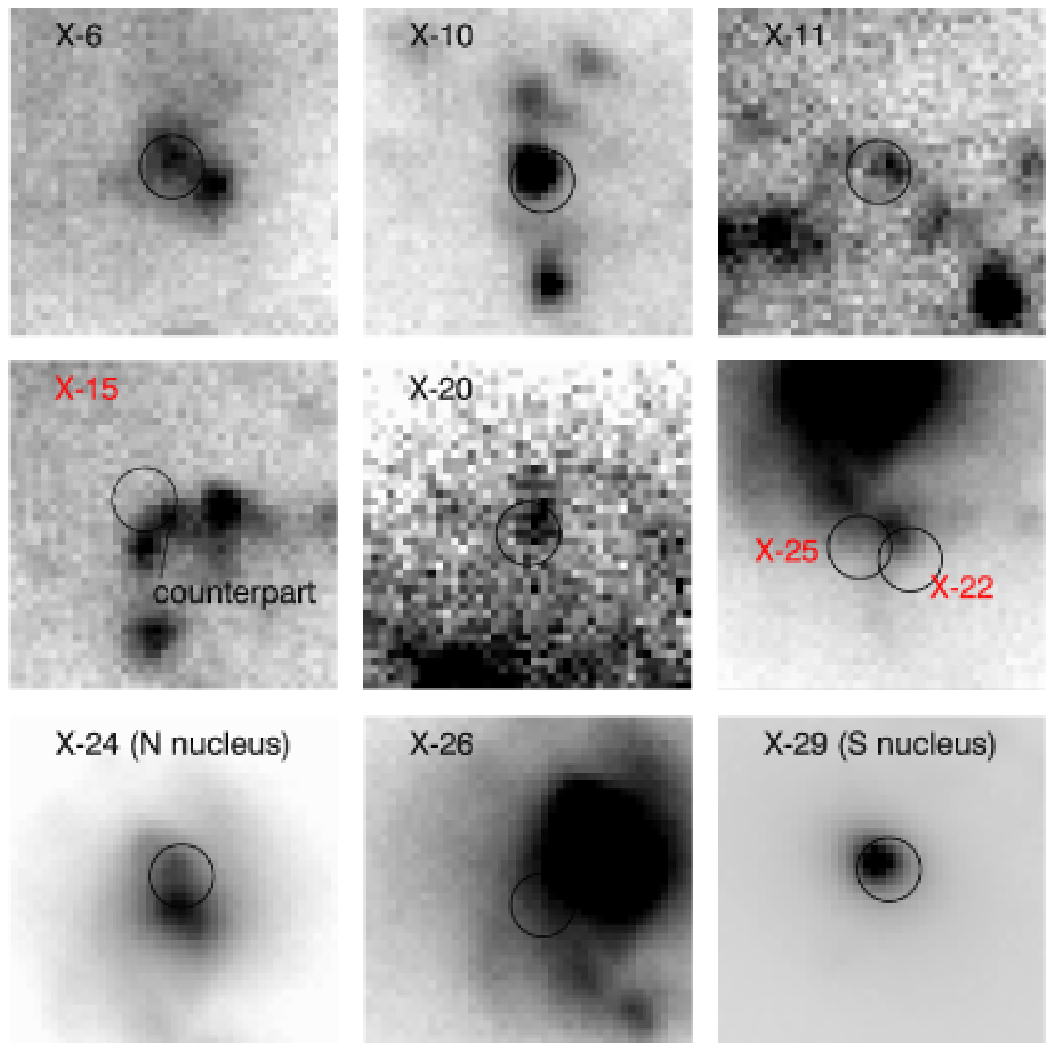}{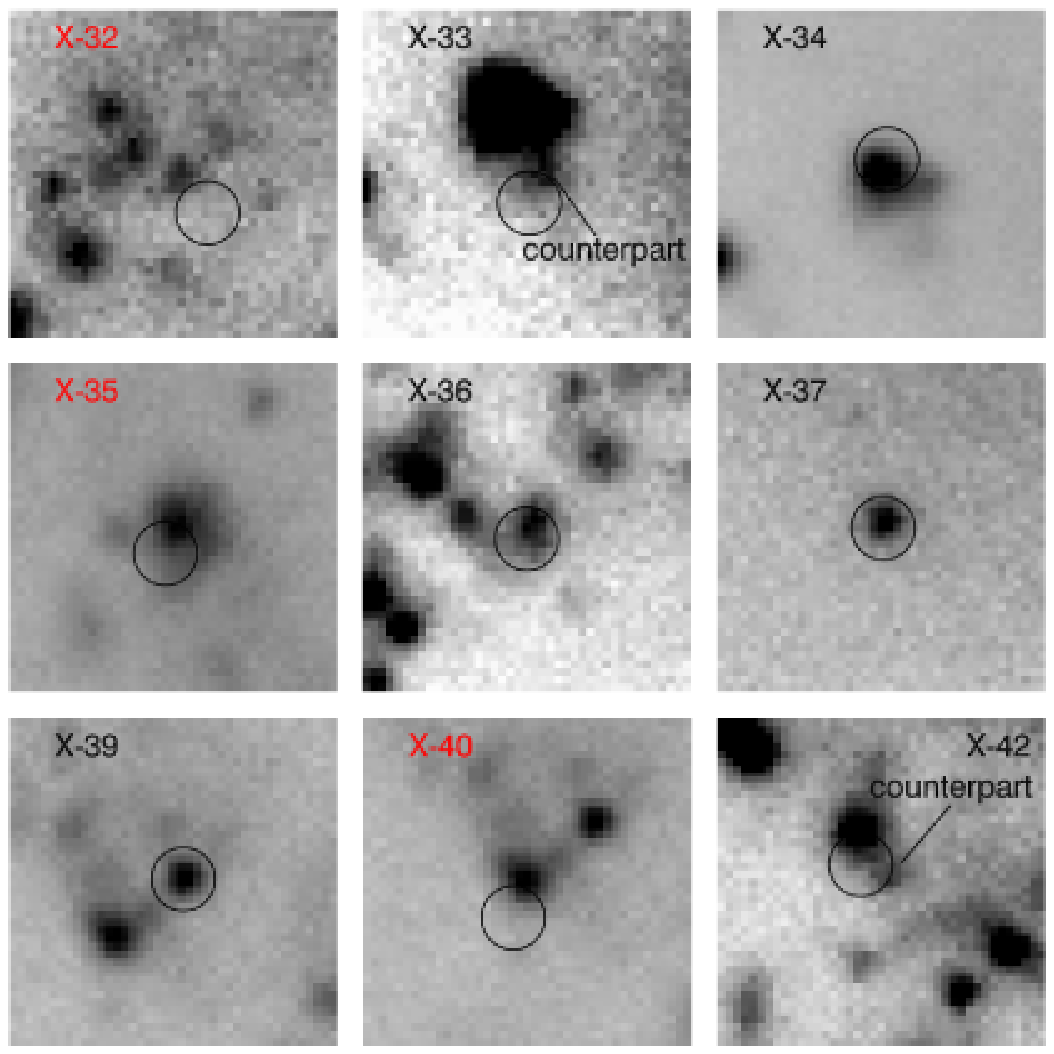}
\figurenum{3}
\caption{Subimages highlighting X-ray positions for all X-rays sources
with IR counterparts.  Positional error circles are $1\farcs0$ in
radius.  Image field-of-view is $10\farcs0\times10\farcs0$ and north
is up.  We label ``strong'' matches in black and ``possible'' matches
in red.
\label{Fig.3}}
\end{figure}

\clearpage

\begin{figure}
\figurenum{4a}
\plotone{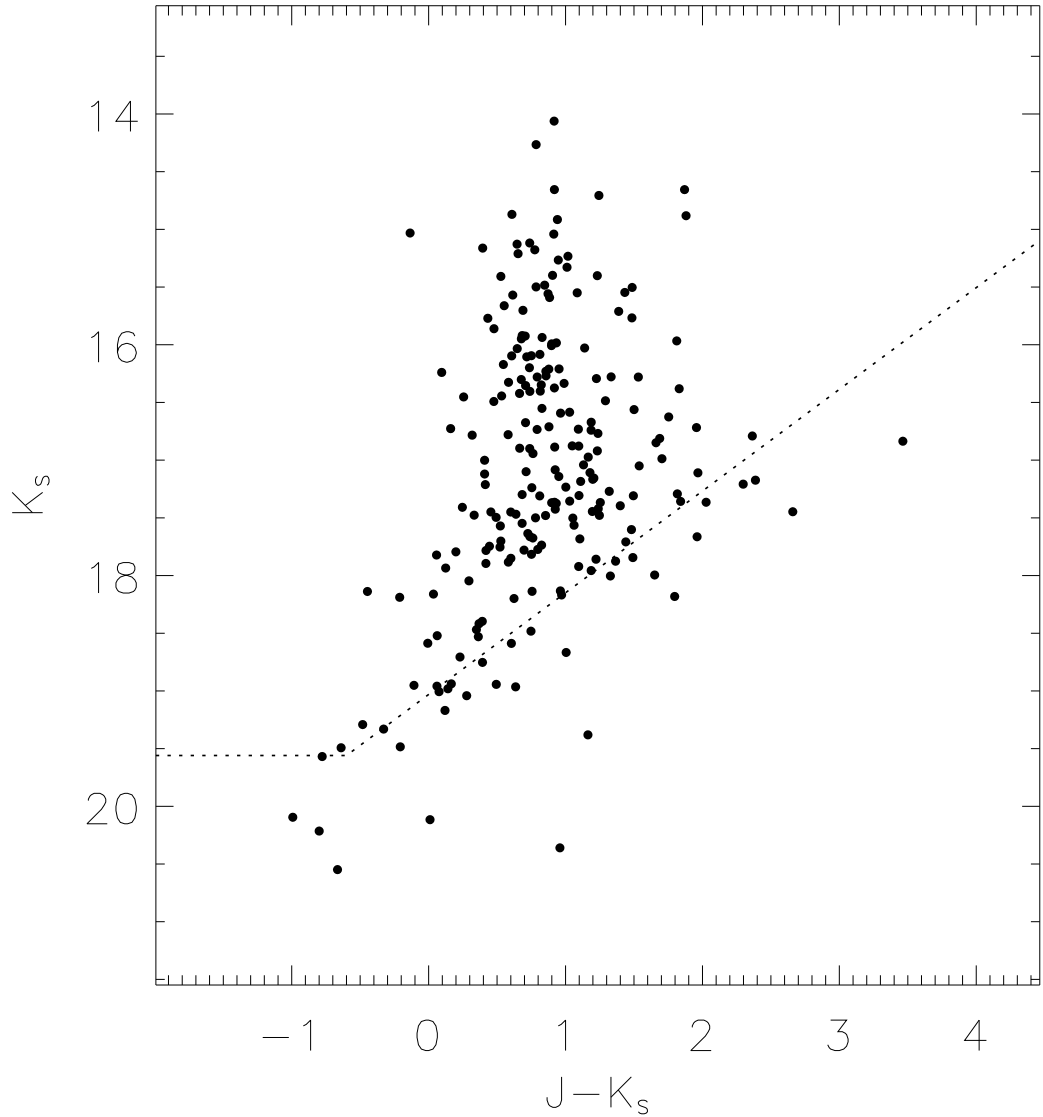}
\caption{(a) $(J-K_s)$ vs. $K_s$ CMD for all clusters with photometry.
In each of the next three plots, the dotted line shows our cutoff for
statistical analyzes. (b) Here clusters with X-ray counterparts are
designated.  Note that one IR counterpart falls below our sensitivity
limit.  Since its $K_s$ magnitude is still above our $K_s$ cutoff, we
retained this source in our analysis.  (c) These clusters are broken
down by their luminosity class.  LLX: $L_X$ $<$ 3$\times10^{38}$ ergs
s$^{-1}$, HLX: 3$\times10^{38}$ ergs s$^{-1}$ $<$ $L_X$ $<$
1$\times$$10^{39}$ergs s$^{-1}$, and ULX: $L_X$ $>$ 1$\times$$10^{39}$
ergs s$^{-1}$.
\label{Fig.4a}}
\end{figure}

\clearpage

\begin{figure}
\figurenum{4b}
\plotone{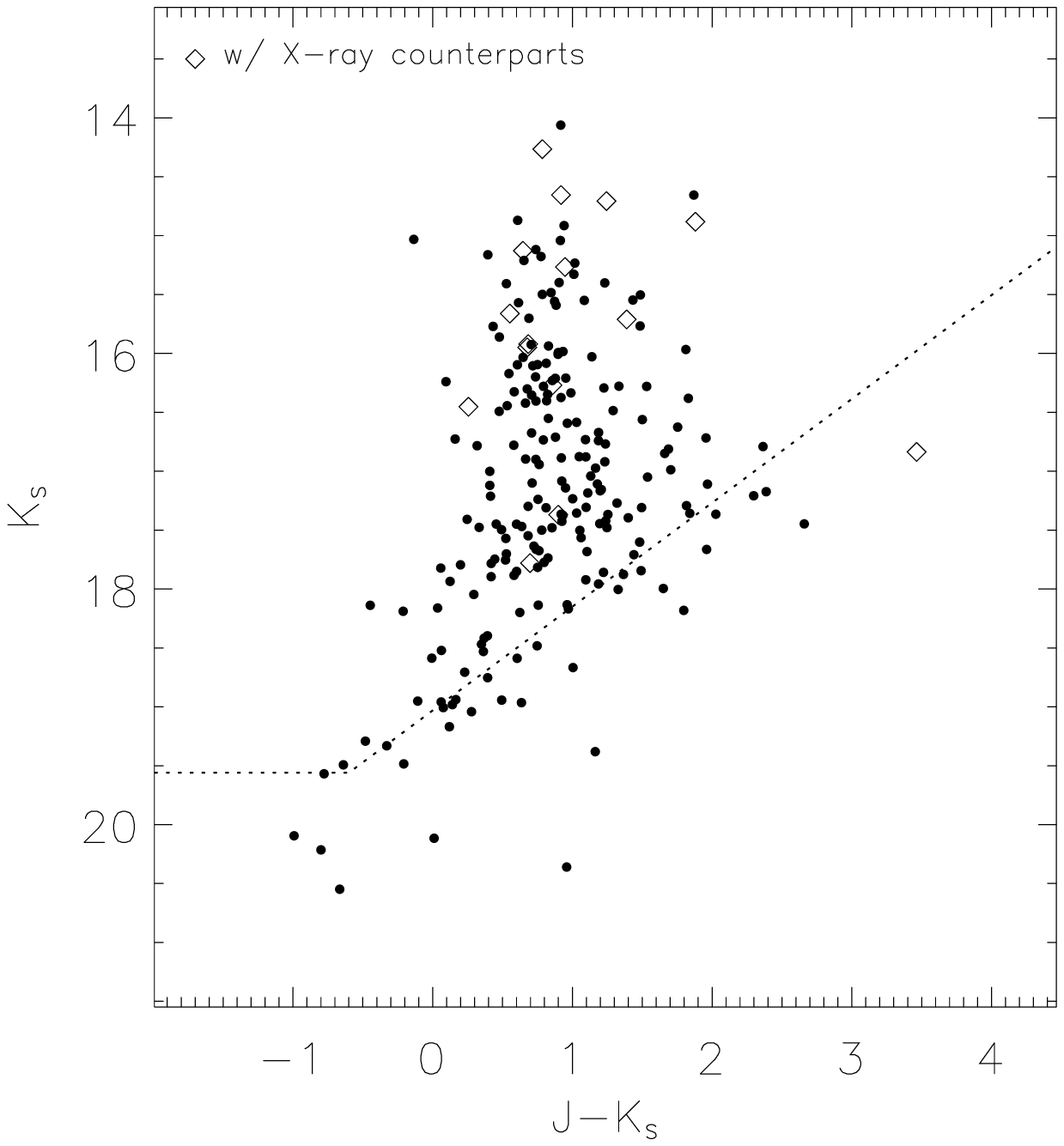}
\caption{\label{Fig.4b}}
\end{figure}

\clearpage

\begin{figure}
\figurenum{4c}
\plotone{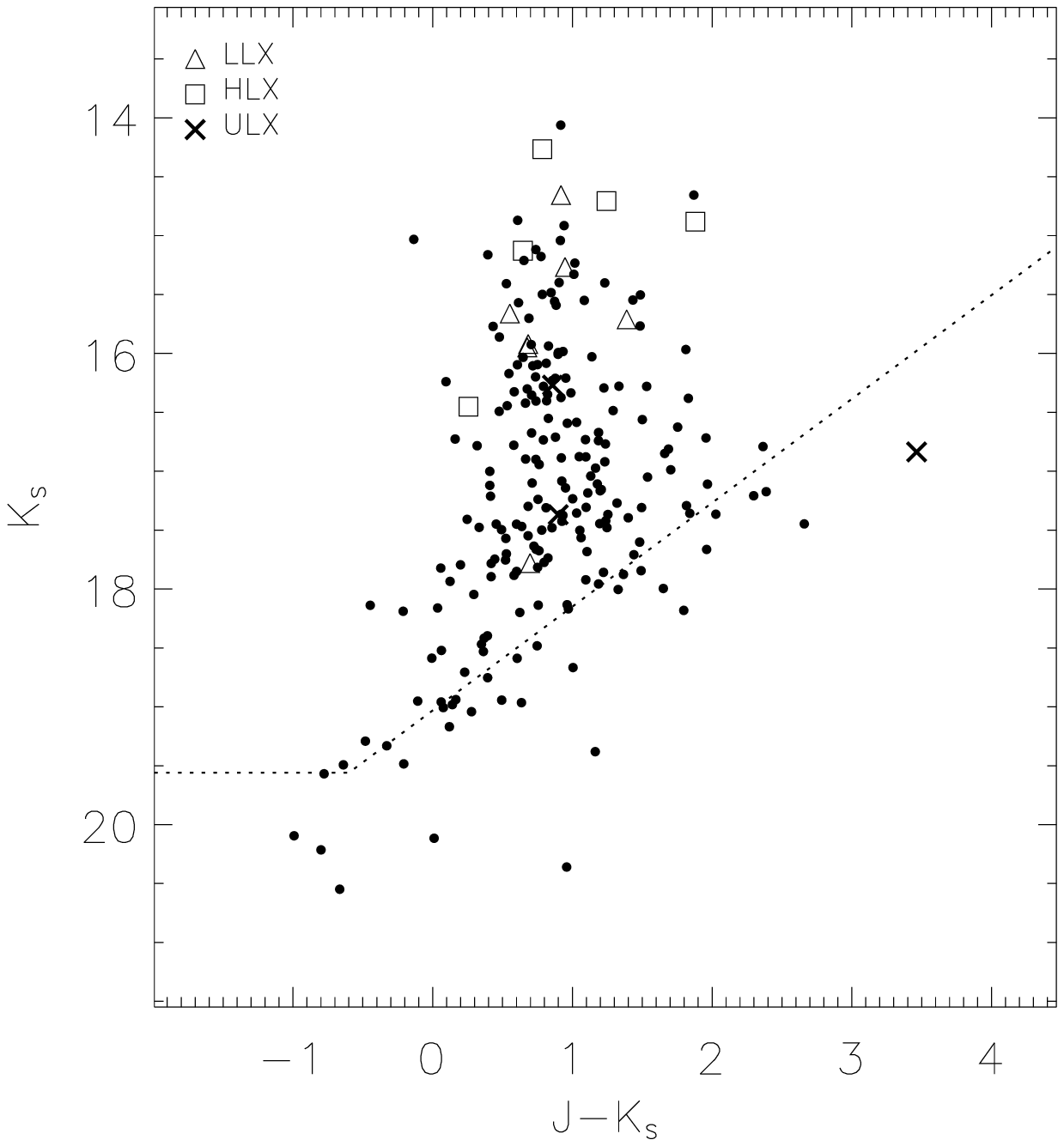}
\caption{\label{Fig.4c}}
\end{figure}

\clearpage

\begin{figure}
\figurenum{5}
\plotone{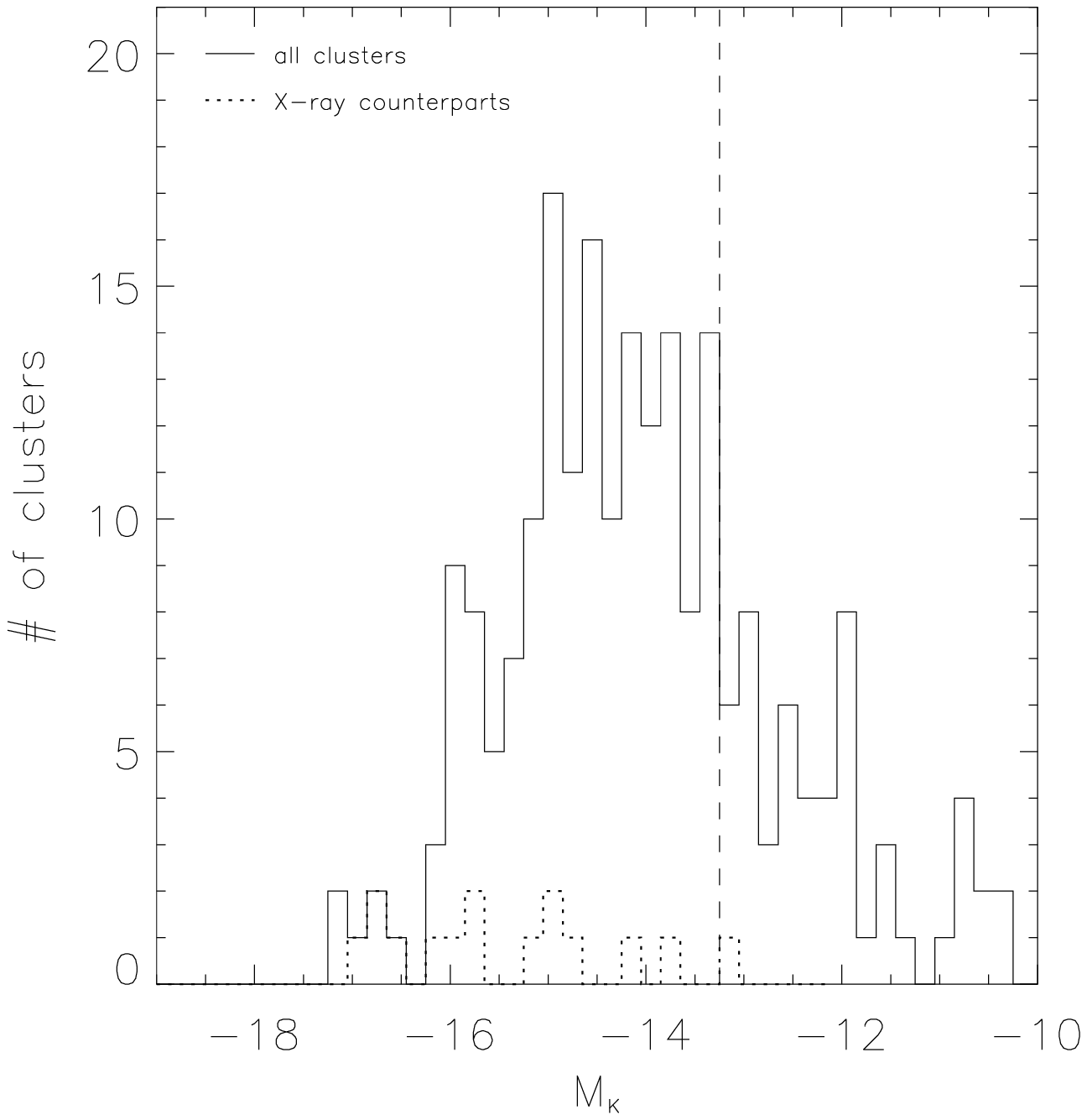}
\caption{$M_{K_s}$ histogram for all clusters.  Binned by 0.2
magnitudes.  Dashed line shows cutoff for statistical analyzes.
\label{Fig.5}}
\end{figure}

\clearpage

\begin{figure}
\figurenum{6}
\plotone{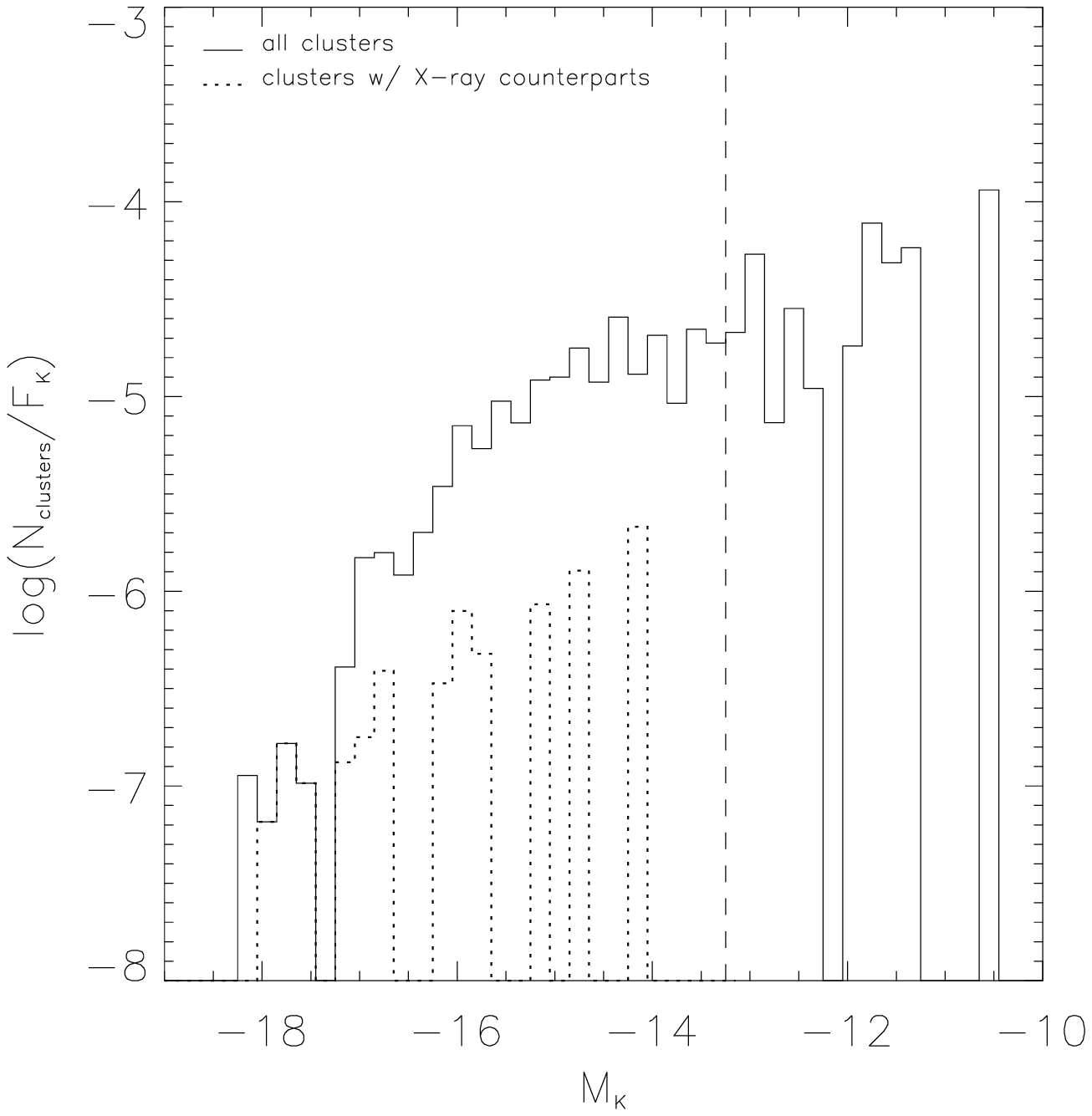}
\caption{This figure is the same as Figure 5, except that we have
plotted $M_{K_s}$ versus the number of clusters per bin divided by the
mean flux in that bin.  Arguing that mass is proportional to
flux, this graph shows the probability of finding a cluster with a
given mass. 
\label{Fig.6}}
\end{figure}

\clearpage

\begin{figure}
\figurenum{7}
\plotone{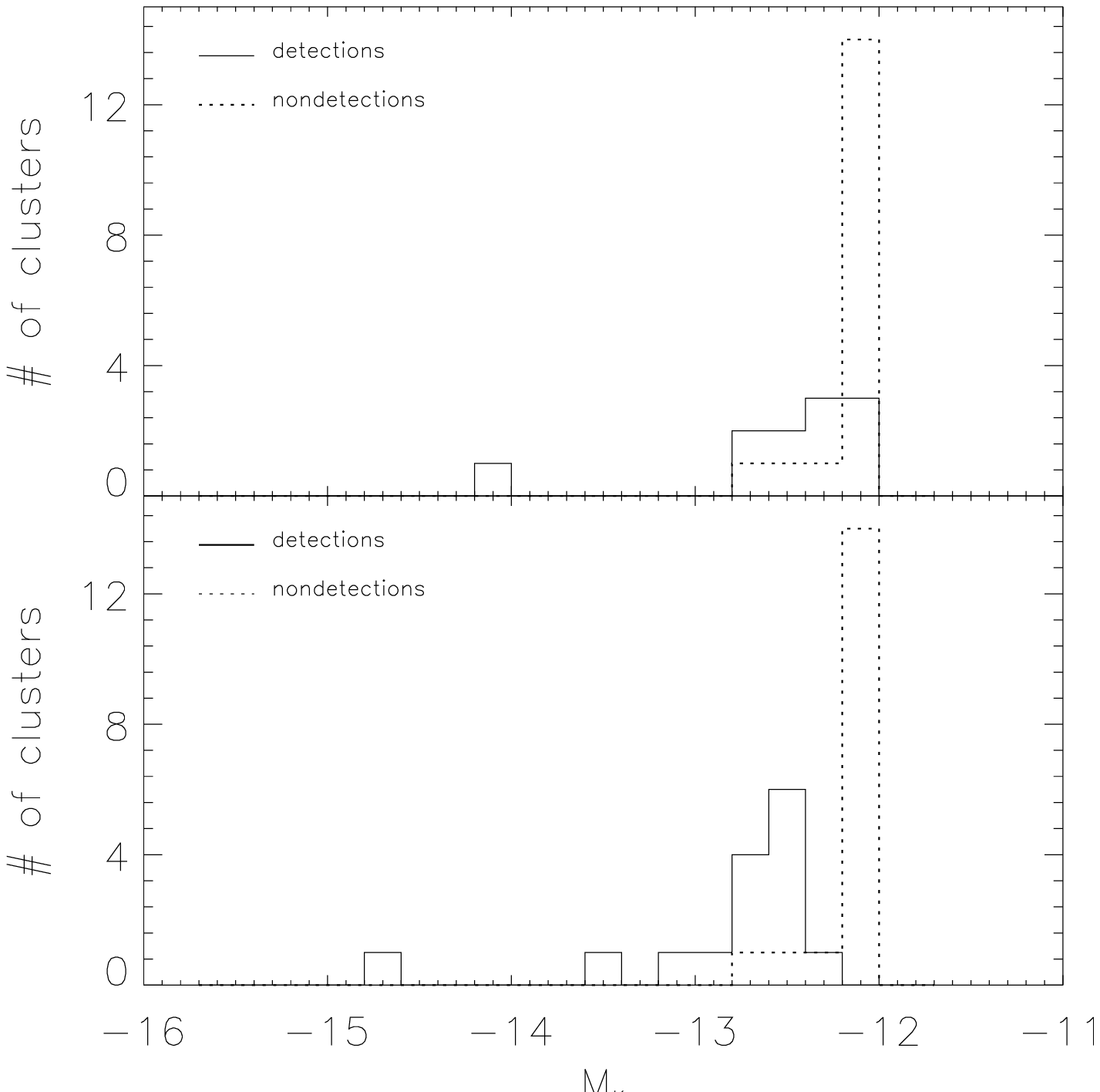}
\caption{Theoretical, $M_{K(lim)}$ histograms for all X-ray sources
with counterparts, detections, and those without, nondetections, using
bins of 0.2 magnitudes.  For detections, reddening computed using
$N_{H}(PL)$ (top) and ``color method'' (bottom).
\label{Fig.7}}
\end{figure}

\end{document}